\documentclass[a4paper, amsfonts, amssymb, amsmath, reprint, showkeys, nofootinbib, twoside]{revtex4-1}
\usepackage[english]{babel}
\usepackage[colorinlistoftodos, color=green!40, prependcaption]{todonotes}
\usepackage[utf8]{inputenc}
\usepackage{amsthm}
\usepackage{mathtools}
\usepackage{physics}
\usepackage{xcolor}
\usepackage{graphicx}
\usepackage[left=23mm,right=13mm,top=35mm,columnsep=15pt]{geometry} 
\usepackage{adjustbox}
\usepackage{placeins}
\usepackage[T1]{fontenc}
\usepackage{lipsum}
\usepackage{csquotes}
\usepackage[pdftex, pdftitle={Article}, pdfauthor={Author}]{hyperref} 
\begin{document}
\title{Controlling phase separation in microgel-polymeric micelle mixtures using variable quench rates}

\author{S L Fussell,$^{\ast}$\textit{$^{123}$} C P Royall,\textit{$^{1234}$} and J S van Duijneveldt\textit{$^{a}$}}
    \email[Correspondence email address: ]{sian.fussell@bristol.ac.uk}
    \affiliation{\textit{$^{1}$School of Chemistry, University of Bristol, Cantock's Close, Bristol, BS8 1TS, UK.}}
    \affiliation{\textit{$^{2}$~HH Wills Physics Laboratory, University of Bristol, Tyndall Avenue, Bristol, BS8 1TL, UK.}}
    \affiliation{\textit{$^{3}$~Bristol Centre for Functional Nanomaterials, University of Bristol, Tyndall Avenue, Bristol, BS8 1TL, UK.}}
    \affiliation{\textit{$^{4}$~Gulliver UMR CNRS 7083, ESPCI Paris, Universit\'e PSL, 75005 Paris, France.}}

\date{\today} 

\begin{abstract}
We investigate the temperature dependent phase behaviour of mixtures of poly(N-isopropylacrylamide) (pNIPAM) microgel colloids and a triblock copolymer (PEO-PPO-PEO) surfactant, which undergoes micellisation. Gelation in these systems results from an \emph{increase} in temperature. Here we alter the heating rate and observe that the mechanism for aggregation changes from one of depletion of the microgels by the micelles at low temperatures to association of the two species at high temperatures. We use this to access multiple structures at one microgel concentration. Samples with a micelle concentration above 7 wt\% were found to demix into phases rich and poor in microgel particles at temperatures below 33$^{\circ}$C, under conditions where the microgels particles are partially swollen. Under rapid heating full demixing is bypassed and gel networks are formed instead. The temperature history of the sample therefore allows for kinetic selection between different final structures, which may be metastable.
\end{abstract}


\maketitle

\section{Introduction} \label{sec:outline}

Phase separation from a one-phase fluid to a demixed system can be achieved very differently in atomic and molecular systems on the one hand and soft matter on the other. In an atomic or molecular system, the interactions between the constituents themselves are largely independent of the temperature, and a quench would entail a lowering of temperature. In soft matter systems, on the other hand, one may often \emph{integrate out} the degrees of freedom of smaller components, and consider \emph{effective} interactions between larger species such as colloidal particles \cite{Likos2000}. In some such systems, for example colloid-polymer mixtures \cite{Poon2002}, the concentration of one species (here the polymer) may play the role of an (inverse) \emph{effective temperature}. Thus here effectively ``quenching'' the system pertains to increasing the polymer concentration  \cite{Poon2002}. 

While much research is carried out at room temperature, one may enquire as to the response of soft materials to changes in temperature. These are often complex. In the case of colloid polymer-mixtures for example, upon heating, polymer molecules can expand, thus increasing the polymer volume fraction and hence effectively quenching the system. In this, one may observe the counter intuitive behaviour of (effectively) quenching by heating and raising the effective temperature by cooling \cite{Taylor2012}. Often more complex behaviour can be observed, due for example to changes in hydrophobicity as a function of temperature between different species in the system. The system of interest here is a mixture of microgel particles and triblock copolymer micelles, which exhibits a complex response to temperature, in addition to a composition dependent phase behaviour typical of soft materials \cite{fussell2019,Gottwald2004,Vilikov2002,Camerin2020,Mohanty2017}. Systems that are capable of triggered gelation or viscosity-switchable materials are sort after for reconstructive surgery and drug delivery applications \cite{Sierra-Martin2012}.

In the case that the interactions are controlled through the composition of the system (rather than temperature), the attractions that drive this phase separation can have a variety of sources, including the depletion interaction induced by \emph{non-absorbing} polymers \cite{asakura1954,Poon2002}, the attractions can come from \emph{bridging} between colloidal particles \cite{Zhao2012b,gao2015b} or through the critical Casimir effect \cite{Shelke2013, Gnan2014}. Under the right conditions such attractions can lead to a network formed through spinodal decomposition, which undergoes arrest due to the asymmetry in viscosity of the two demixing phases, which is termed viscoelastic phase separation, or sometimes no attractions are actually needed to form such a network of fluids with very different viscosities \cite{Ferreiro-Cordova2020}. Whatever the origins of the emergence of the viscoelastic network, \emph{understanding} colloidal gelation due to arrested spinodal decomposition is a long-standing challenge  \cite{Poon2002,Lu2008,Zaccarelli2007,Cipelletti2005,Verhaegh1997,Truzzolillo2014}, although the mechanism of solidification has been related to the emergence of rigid clusters, leading to a gel with viscoelastic mechanical properties \cite{Royall2008, Royall2015}.

Gels are far-from-equilibrium systems and their properties exhibit complex time-dependent behaviour, as they relax towards the equlibrium state of phase coexistence between the colloid-rich (high-viscosity) phase and colloid-poor (low viscosity) phase \cite{Cipelletti2005,Bonn2017,Royall2012,Bartlett2012,Aime2018,Harich2016,
Tsurusawa2017a,zhang2013,Griffiths2013, Royall2015}.  Studies of assembly of so-called ``sticky spheres'' out of equilibrium, which is a basic model system which undergoes gelation where the colloid-rich phase may crystallise  \cite{klotsa2011}, have shown that assembly at different temperatures leads to a complex response of fast yet poor-quality assembly upon deep (effective) quenches, and slower but high-quality assembly when the (effective) temperature is higher \cite{whitelam2015}. This is backed up by experimental work where the system evolved more quickly to its equilibrium demixed state in the case of weak quenches \cite{zhang2013,razali2017}. Time-dependent assembly protocols have also been investigated, and it has been found that slow quenches promote high-quality assembly \cite{Royall2012}, and time-dependent protocols of a deep quench at short times followed by a shallow quench at longer times optimise the process of assembly further \cite{Klotsa2013}.
Here we consider a soft matter system that displays both liquid-liquid phase separation and formation of gel networks, depending on the heating rate.

Our model system is a mixture of microgels and micelle-forming triblock copolymers which exhibits a complex response to temperature  \cite{fussell2019}.
Microgels are colloidal particles comprised of a cross linked polymer  \cite{Schneider2018,Wedel2016,Varga2001a,Bonham2014,Bergman2018,Pelton1986} which exhibit some similar phenomena to hard colloids \cite{yunker2014,royall2013,Bergman2018}. In particular, poly(N-isopropylacrylamide) (pNIPAM) microgels are known to form colloidal 
gels in the presence of additives, such as salt, which screens the electrostatic interactions \cite{Hu2004,Immink2019,Bischofberger2015}. Their swelling behaviour is temperature responsive, which is a result of the polymer-solvent interactions \cite{Nigro2015}. In our system, a number of phenomena occur as the temperature is increased between $25^\circ$C and $40^\circ$C. pNIPAM microgels exhibit a so-called volume phase transition where the particles which are swollen at room temperature undergo a collapse \cite{McPhee1993}. The micelles are formed from a PEO-PPO-PEO triblock copolymer \cite{Alexandridis1994}. 

Upon heating, there are two underlying phenomena: firstly, the microgels undergo the volume phase transition and collapse over a temperature range of around $10^{\circ}$C, between $30$ and $40^{\circ}$C. \cite{Bandyopadhyay2017} The degree of swelling is controlled by the cross link density in the particle, in our system this volume phase transition corresponds to a two-fold change in diameter (i.e. 8-fold change in volume). It is important to note that this does not correspond to a \emph{complete} collapse and expulsion of all solvent, rather, about half of the material within the microgel is solvent in the collapsed state \cite{sbeih2019}. Moreover, the distribution of the pNIPAM and solvent is non-uniform with a higher density of pNIPAM and cross links towards the centre and a lower density of polymer and cross links towards the periphery of the particle \cite{Ninarello2019}. The volume phase transition which thus provides the ability to tune the interaction potential between pNIPAM microgels with temperature \cite{Sierra-Martin2012}. When the microgels are swollen, they are purely repulsive \cite{Heyes2009a, Mansson2019}, due to the minimal van der Waals (vdW) interactions. With increasing temperature, the deswollen particles are often modelled as ``sticky spheres'', due to increased vdWs interactions and hydrophobic interactions \cite{Bergman2018}. 

\begin{figure}[h]
\centering
  \includegraphics[width=\columnwidth]{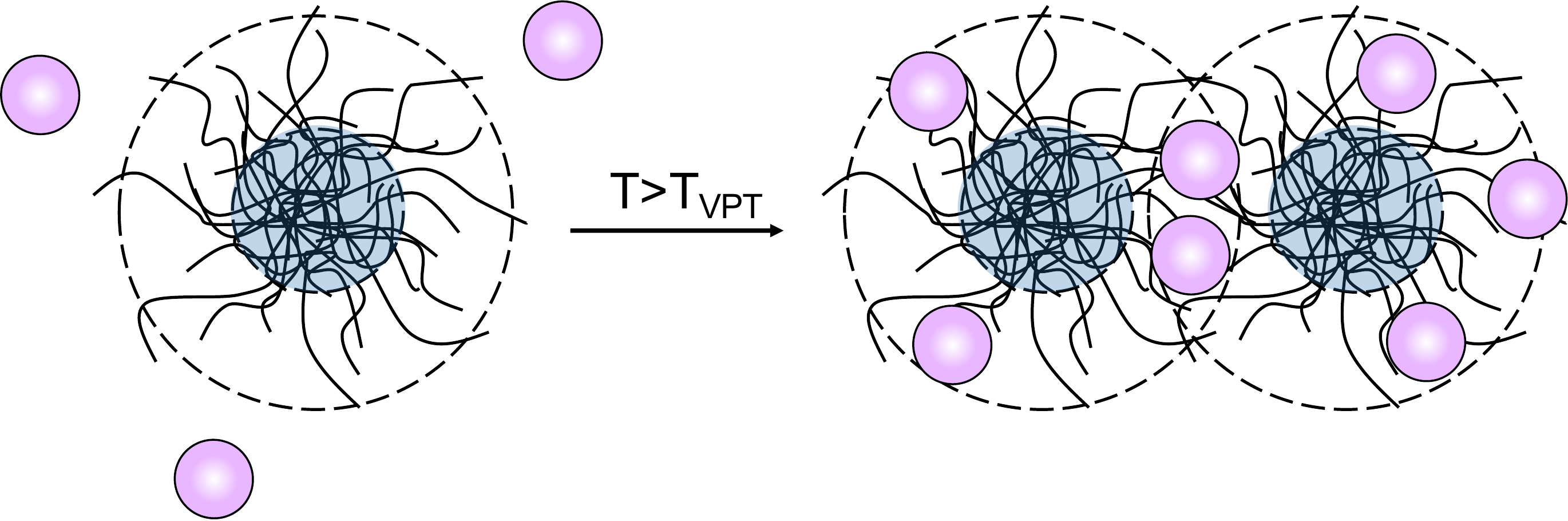}
  \caption{Scheme representing the gelation mechanism between pNIPAM microgels and triblock-copolymer micelles at temperatures above the $T_{\text{VPT}}$. The micelles are represented by the pink circles. \cite{fussell2019}}
  \label{fgr:initial}
\end{figure}

The second effect, which drives the gelation in our system \cite{fussell2019}, is that upon heating above the volume phase transition temperature, the micelles associate with the microgels due to hydrophobic interactions, which leads to the microgel particles aggregating and undergoing gelation (Fig. \ref{fgr:initial}). Thus, here, the system is \emph{effectively} quenched by \emph{increasing temperature} \cite{fussell2019}, in a manner reminiscent of the mixtures of colloids and non-absorbing polymers noted above \cite{Taylor2012}. Here, we explore the response of our system to \emph{time-dependent temperature protocols}. We find a complex behaviour resulting from the competition between thermodynamic forces driving phase separation and slow dynamics. By changing the \emph{effective} quench rate, we tune the system towards arrested phase separation, when the rate of temperature change is fast with respect to the demixing kinetics or towards phase separation when the rate is slow with respect to demixing kinetics. 


We use differential interference contrast and confocal microscopy to identify the structure of the gel networks, and study the phase behaviour of this system. Firstly, we observed the high temperature gelation of these species, but also a change in the phase behaviour, as a result of altering the conditions by which gelation occurred. Secondly, under specific conditions we observe demixing (phase separation), where microgel rich and poor liquid regions form. We highlight the trends in phase behaviour that results from tuning the gelation conditions (quench rate) and the sample concentration.  

This paper is organised as follows: in section 2 we show our methods, in section \ref{sectionResults}, we present our main findings including the phase behaviour, control over phase using quench rates and detailed characteristics of our system, including structure and droplet growth kinetics.  
\section{Methods} \label{sec:develop}
  
\subsection{Materials}

For the microgel synthesis N,N'-methylenebis(acrylamide) (99\% Sigma Aldrich), N-isopropylacrylamide (99\% Acros Organics), potassium persulfate (>99\% Sigma Aldrich), sodium dodecylbenzenesulfonate (technical grade Sigma Aldrich), fluorescein isothiocyanate (98\% Sigma Aldrich) and methacryloxyethyl thiocarbonyl rhodamine B (Polysciences) were used as received without further purification. The triblock-copolymer surfactant was labelled using Nile red dye (technical grade Sigma Aldrich). Synperonic SE/P105 is the triblock-copolymer used for all the experiments, a polyethyleneoxide-polypropyleneoxide-polyethyleneoxide (PEO-PPO-PEO) type polymer (6500 MW). 

\begin{figure*}
\centering
  \includegraphics[height=7cm]{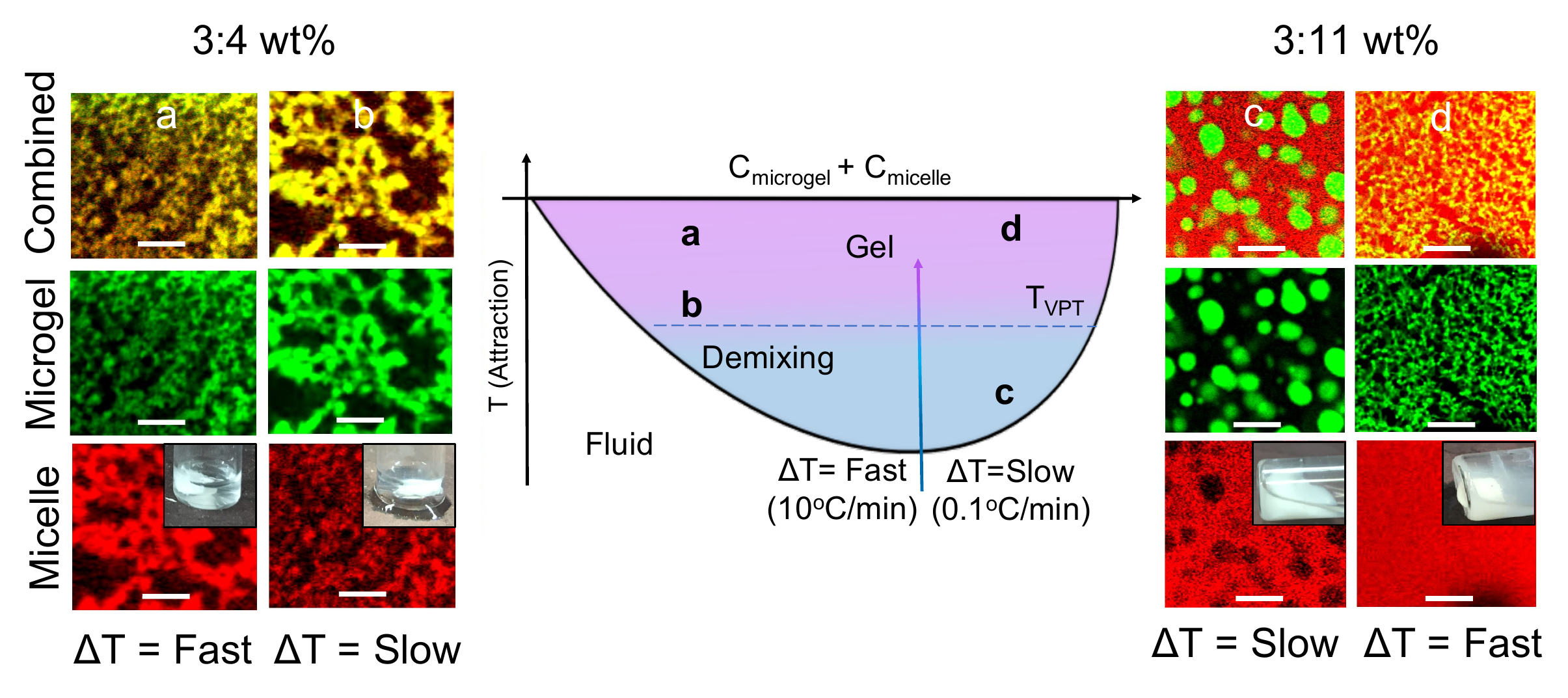}
  \caption{Scheme representing the observed behaviour of samples of pNIPAM and triblock-copolymer surfactant, imaged using confocal microscopy. For a sample heated slowly, phase separation is observed (blue) at lower temperatures and gelation (pink) at higher temperatures. If the sample is heated quickly, the liquid-liquid phase separation is bypassed, and only gels (pink) are observed at all concentrations. Images highlighting the difference in both the macroscopic and microscopic phase behaviour of pNIPAM samples are included. All samples contain 3 wt\% pNIPAM, the images on the left are of samples with 4 wt\% triblock-copolymer. The images on the right are of samples with 11 wt\% triblock-copolymer. Images on the bottom have been heated up slowly in a oven set to 40$^{\circ}$C (b,c). Images on the top are of samples which were submerged into a hot water bath (a,d). Both images of the macroscopic gel and the structure of the gel observed using confocal microscopy have been included. The green images are of the labelled microgels, the red images are of the labelled micelles and the combined image of both dyes  is also included. The scale bar included for all the confocal images is 10 $\mu$m. The corresponding location of each sample on the phase diagram has been highlighted. The $T_{\text{VPT}}$ of pNIPAM has been highlighted on the figure using a dashed blue line.}
  \label{fgr:phase}
\end{figure*}

\subsection{pNIPAM microgel synthesis}

The microgels were synthesised using precipitation polymerisation.\cite{McPhee1993} Each synthesis was conducted in a 1000 mL three necked round bottom flask, fitted with a condenser, an argon inlet and an overhead stirrer. The NIPAM (12.5 g), N,N'-methylenebis(acrylamide) (1.0 g) and sodium dodecyl benzenesulfonate (0.5 g) were added to the flask along with 475 mL of deionised water and heated to 70$^{\circ}$C with stirring under an inert atmosphere. Once heated potassium persulfate (0.5 g dissolved in 25 mL of DI water) was slowly added to the reaction and the solution left stirring at temperature for 4 hours. The microgels produced were then purified using dialysis against deionised water for 2 weeks. Fluorescein labelled particles were synthesised using 9 mg of fluorescein isothiocyanate and 15 $\mu$L 3-amino-propene, added to 38 mL of a solution of 10$^{-4}$ M sodium hydroxide. This solution was then added at the start of the pNIPAM reaction discussed above, in place of 38 mL of water.\cite{Meijer2015} All particles were then characterised using a Malvern Autosizer 4800 with a 532 nm laser at a 90$^{\circ}$ scattering angle. This was used to determine the diameter of the particles and to study the deswelling behaviour. All samples were diluted using deionised water and left to equilibrate at each temperature for 15 minutes. 

\subsection{Microscopy}

An Olympus BX-51 DIC microscope was used to obtain the differential interference contrast (DIC) images. All images were captured using a Pixelink 5MP colour CCD PL-B625CU camera and a 20$\times$ objective. A Linkam PE120 heating stage was used to control the slide temperature. The samples were prepared by adding a drop of solution to the centre of a doughnut shaped sticker (Office Depot) attached to a microscope slide, and a cover slip was then placed onto the sample. This gave a sample environment for imaging with a of diameter 5 mm and a depth of 200 $\mu$m. For the fast heating experiments, the samples were heated at a rate of 10$^{\circ}$Cmin$^{-1}$ from room temperature to 35$^{\circ}$C. For the slow heating experiments, the samples were heated at a rate of 0.1$^{\circ}$Cmin$^{-1}$ to a specified temperature above the gelation temperature. For all experiments this was 35$^{\circ}$C unless otherwise stated. The temperature quoted as the gelation temperature is the temperature the microscope stage read when gelation occurred. There may be some discrepancies between the temperature of the slide and the read out temperature, these are assumed to be equivalent across all samples therefore the general trends should remain the same. 

Confocal images were taken using a Leica SP8 confocal microscope. The sample preparation was as described for DIC microscopy along with the heating protocol. The micelles were labelled with Nile red, a 543 nm HeNe laser was used to excite the dye. The microgels were labelled with fluorescein, an Argon laser (488 nm) was used to excite the dye. All samples were heated using a TC-1-100s temperature controller, along with a Bioscience Tools objective heater, set to 35$^{\circ}$C.

The DIC images have been edited using ImageJ software. The images have been converted to grey scale, then to the `glow' colour table, then the image inverted. This was done to highlight the polymer rich phase using this imaging technique. An example of the process to edit the DIC images is included in the ESI (Fig. S1 ESI).

\subsection{Macroscopic characterisation of phase behaviour}

To replicate the fast heating experiments for macroscopic gels, the samples were made up in sealed vials and submerged into hot water (approximately 70$^{\circ}$C). The phase behaviour was then recorded after 10 minutes. To replicate the slow heating experiments, samples were placed into a oven set to 40$^{\circ}$C and allowed to slowly heat up from room temperature, and the phase behaviour was recorded after 24 hours.

\subsection{Image analysis}

To determine the size of the droplets formed, the images were fitted manually using Fiji image analysis software. The diameter of 100 droplets was measured and averaged to obtain an estimate of diameter and the polydispersity in the sample. To investigate the droplet growth over time, the data obtained from image analysis was plotted. The droplet size as a function of time was plotted on a log-log plot and the linear fit was used to obtain the power law exponent. The data was fitted up to the point where the droplet growth plateaus, which varied depending on the sample.

\section{Results and discussion} \label{sectionResults}

In this section, we present the temperature induced gelation of pNIPAM microgels and triblock-copolymer micelles. We present the temperature response of these systems characterised at both the macroscopic (cm) and microscopic ($\mu$m) length scale. We characterise the structure formed in these mixtures using DIC and confocal microscopy. We also use DIC imaging to characterise the temperature where the microgel particles
aggregate (Fig. \ref{fgr:phase}), and highlight how the temperature response at the particle level can be related to the structures observed. We alter the heating rate in order to determine the relationship between heating rate and the structures observed. We also use image analysis to characterise the droplet phase formed under appropriate conditions. We use confocal microscopy to identify the position of species in the structures. 
We note that the critical micelle concentration for the triblock-copolymer surfactant is 0.3 wt\% at 25$^{\circ}$C and 0.005 wt\% at 35$^{\circ}$C, therefore it is assumed in this work that the majority 
added is present in the form of micelles, rather than as free surfactant \cite{Alexandridis1994}.

\subsection{Heating rate}

\begin{figure*}
\centering
  \includegraphics[height=5cm]{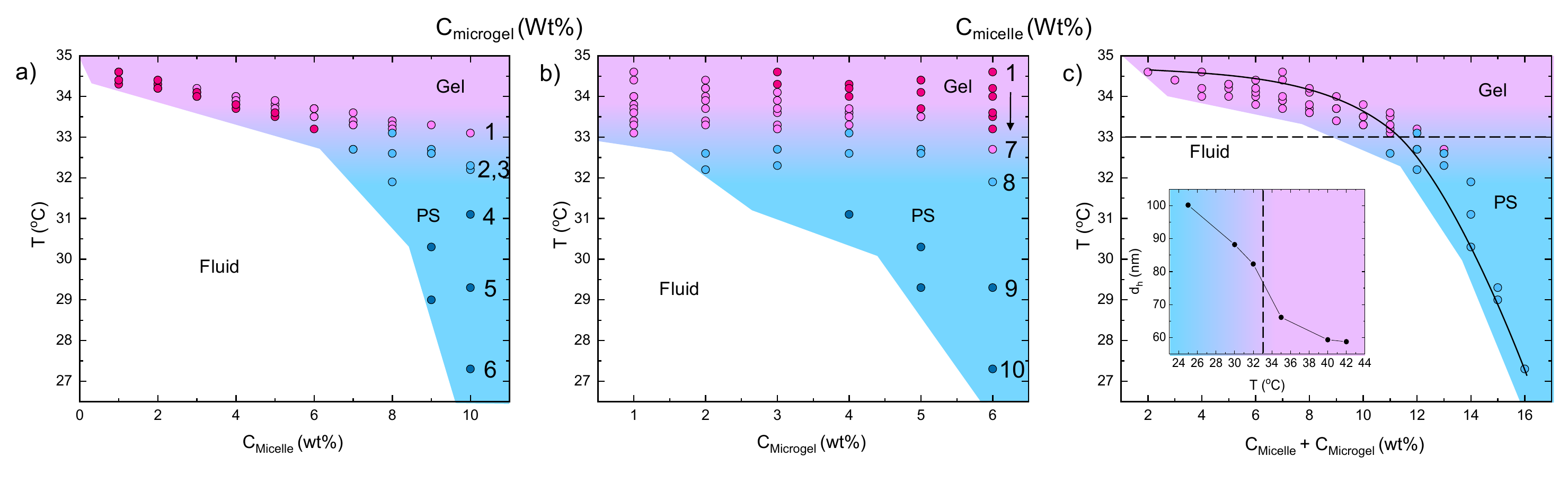}
  \caption{Summary of the aggregation temperatures observed using DIC microscopy. The samples were heated at a rate of 0.1$^{\circ}$Cmin$^{-1}$, and the temperature was recorded as the point where aggregation was first observed. The colours represent the phase observed on the microscope. Pink points represent concentrations where gels were observed, blue points indicate concentrations where phase separation was observed. The black solid curve has been added to highlight the general trend of decreasing gelation temperature with increased triblock-copolymer concentration. If gelation occurs at temperatures above about 33$^{\circ}$C, then gels are always observed. The deswelling curve for pNIPAM microgels has also been included (c), to highlight how at 33$^{\circ}$C the particles are still somewhat swollen, this temperature is again indicated by a dashed line. The shading added is to represent whether gel phases preferentially form (pink) or phase separation (blue) at each temperature/ swelling. The data has been plotted as a function of triblock-copolymer concentration (a), pNIPAM concentration (b) and total polymer concentration (c). The dark pink circles indicate samples that are uniform gels, the light pink circles samples that form gels with voids, light blue circles samples that from droplets and dark blue circles samples that phase separate.}
  \label{fgr:temp2}
\end{figure*}

\begin{figure*}
\centering
  \includegraphics[height=7.5cm]{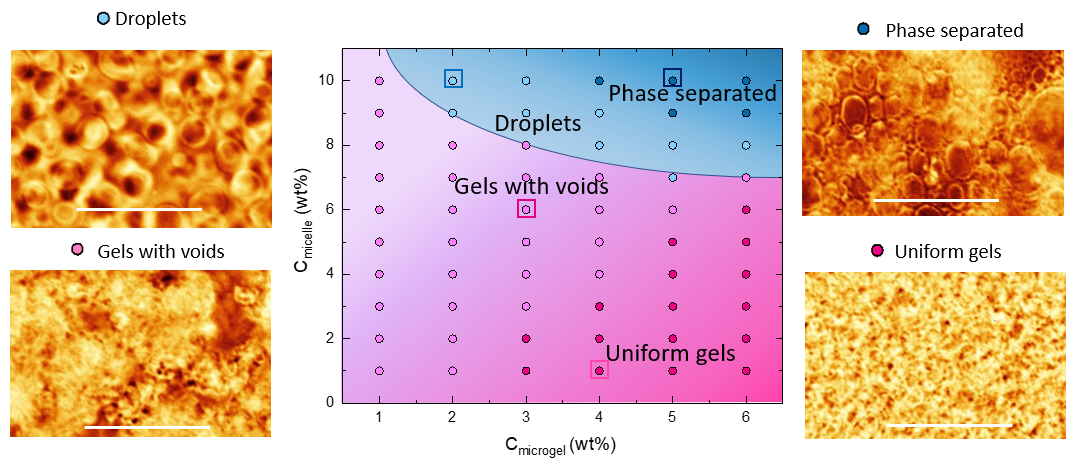}
  \caption{Phase behaviour of mixtures of pNIPAM microgels and triblock-copolymer surfactant, imaged using DIC microscopy. All samples were heated to 35 $^{\circ}$C at a heating rate of 0.1 $^{\circ}$Cmin$^{-1}$ before the final phase was recorded. Pink circles indicate where gel networks were observed. Light pink circles indicate where gels with larger voids were observed, and dark pink circles where uniform gels are observed. Blue circles indicate where phase separation is observed. The light blue circles indicate where isolated droplets were observed, and the dark blue circles indicate where phase separation occurred. The coloured squares indicate the concentrations of the samples in the example DIC images included. The shading has been added to guide the eye. The scale bar on all the images is 40 $\mu$m.}
  \label{fgr:phasediagram}
\end{figure*}

\textit{Fast heating. --- }
We begin our presentation of our results by considering the case of fast heating ($10\ ^\circ$C\ min$^{-1}$ from room temperature to $40\ ^\circ$C).
Samples of pNIPAM microgels and triblock-copolymer surfactant micelles were heated and imaged using DIC microscopy. Upon heating at a rate of 10$\ ^{\circ}$C\ min$^{-1}$, there is very little observable difference in the gel structures, regardless of concentration of \emph{either} of the two species, microgel colloids or micelles. All samples were heated to 35$^{\circ}$C, a temperature above the volume phase transition temperature ($T_{\text{VPT}}$) for pNIPAM, meaning the microgels used in this study are close to fully deswollen (Fig. S2 Electronic Supplementary (ESI)). The structure observed for all samples is a uniform gel network with fine features below the resolution of the DIC microscope. The gel branches are of order 1 $\mu$m width with very few observable pores. An example of the structure observed is included in Fig. \ref{fgr:phasediagram}. There is an increase in the density of the network with an increase in concentration of either microgel colloids or micelles. The mechanism for gelation upon fast heating is one of association, where the microgels and micelles associate due to hydrophobic interactions and hydrogen bonding. This association was characterised in our previous work, using small angle neutron scattering, confocal microscopy and DLS \cite{fussell2019}.

At low microgel concentrations, on experimental timescales, we find that the system remains fluid at elevated temperature, microscopic observation reveals finite sized aggregates. We expect that for sufficient waiting times and system size, gelation would occur \cite{Griffiths2013}. The concentrations investigated here were 1-5 wt\% pNIPAM microgel and 1-10 wt\% triblock-copolymer micelles. The lower concentrations of microgels and micelles show evidence of aggregates forming with heating, as the samples become increasingly opaque compared to microgels alone. At high enough concentrations of the micelles and microgels, the mixtures from macroscopic gels (>3 wt\% microgel), which are originally space spanning and subsequently collapse over time, leaving a liquid supernatant around the dynamically arrested gel \cite{fussell2019}. These gels form reversibly. The mixtures form gels upon heating but redisperse at low temperatures to reform a colloidal suspension. This behaviour has been investigated in our previous work \cite{fussell2019}. 

Confocal microscopy indicates that both the microgels and micelles are present in the gel network, where the green fluorescent signal (microgels) and red fluorescent signal (micelles) can be seen in the same location of the gel. This indicates that association has occurred (Fig. \ref{fgr:phase} a,d). At the highest concentrations of micelles, there is an excess where there is evidence that the micelles are still present in the supernatant. This indicates that the mechanism for gelation is likely through the micelles/surfactants adsorbing into the microgel, preventing the particles from deswelling, resulting in gelation due to the increased vdW interactions between the larger, more hard sphere like microgels. 

\begin{figure*}
\centering
  \includegraphics[height=5.5cm]{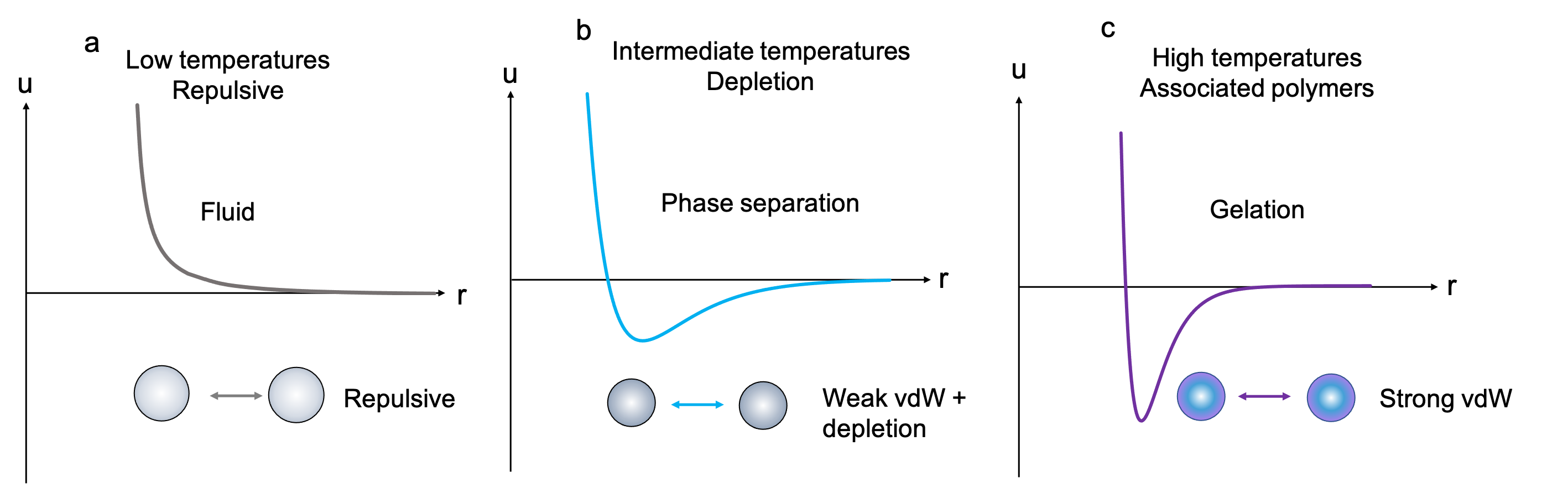}
  \caption{Sketch of the interaction potentials between microgel particles at different temperatures in the presence of triblock-copolymer micelles. At low temperatures (a), the interaction between microgels is purely repulsive (fluid). At intermediate temperatures (b), weak vdW forces now dominate and a depletion force is also present due to the non adsorbing micelles, which results in  liquid phases rich in the two polymer species (phase separation). At high temperatures (c), pNIPAM microgels now associate with the micelles, resulting in the microgels not fully collapsing, there are now strong vdW forces between them and the particles stick together as a result of association (leading to gelation). The blue circles represent the pNIPAM microgels at different extents of deswelling.}
  \label{fgr:interactions}
\end{figure*}

\textit{Slow heating. --- }
Figure \ref{fgr:phase} summarises the general behaviour observed for mixtures of microgels and micelles when samples are heated slowly (0.1$\ ^{\circ}$C min$^{-1}$). The structures observed are totally different compared to those found using fast heating. At high concentrations of micelles and microgels and temperatures below $T_{\text{VPT}}$, demixing occurred at a micelle concentration of around 7 wt\% micelle. Droplets rich in microgels are seen. These droplets contain just the microgel species, with the micelles remaining free in solution. This was evidenced using confocal microscopy, as the fluorescent signal of the micelles and microgels appear in different regions of the image (Fig. \ref{fgr:phase} d). This implies the mechanism for microgel aggregation in this system is one of depletion, where non-adsorbing polymer causes colloidal aggregation. With increasing time or concentration of microgels and micelles, these droplets coalesce leading to phase separation. These demixed phases were observed at temperatures \emph{lower} than the $T_{\text{VPT}}$ for pNIPAM, the temperature where aggregation was originally assumed to occur in these samples. In comparison to structures formed by our system as a result of the fast heating, we see more striking differences at higher microgel and micelle concentrations. At lower concentrations and higher temperatures, gel networks form (a, 4 wt\% micelle), which are similar to those formed at the same composition with the faster heating rate. 

When mixtures of  microgels and micelles are heated quickly (10$\ ^{\circ}$C\ min$^{-1}$), the phase separation is bypassed and colloidal gels are observed at all concentrations (b, d). Whether phase separation or gelation occurs can be related to the quench rate in other colloidal systems. Slow quenching often results in phase separation rather than gelation \cite{Royall2012,Royall2011}. Gels are non-equilibrium structures, and therefore the effects of kinetic trapping can be important. \cite{Poon2002,Lu2008,Zaccarelli2007,Cipelletti2005,Verhaegh1997}. Hence in our system, gel networks are always observed on fast heating (10$^{\circ}$Cmin$^{-1}$) as the equilibrium structure cannot be accessed. For the lower concentration samples, heating rate has little effect on the structure and gel phases are observed at both heating rates (a,b 4 wt\% micelle). However, for the higher concentration samples, large differences in the structure were observed (c,d 11 wt\% micelle). This is where the aggregation of the micelles and microgels occurs at lower temperatures.

\begin{figure*}
\centering
  \includegraphics[height=8cm]{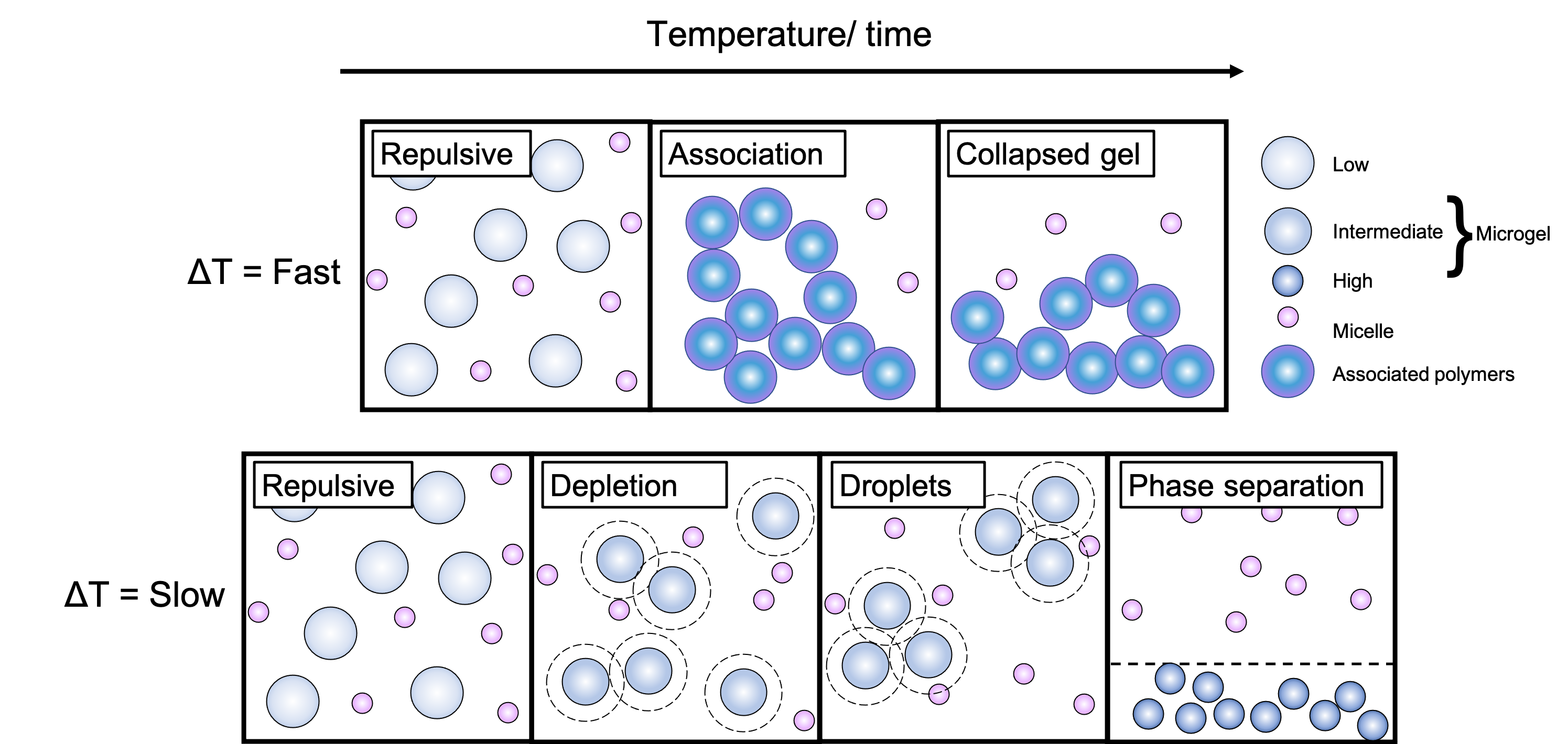}
  \caption{Scheme highlighting the behaviour of pNIPAM microgels in the presence of triblock-copolymer surfactant, comparing the difference when samples are heated slowly and quickly. The interactions are shown as a function of both temperature and time. The top layer shows samples that are heated quickly. At low temperatures the microgel particles are repulsive, then when heated quickly, they associate strongly, likely through the micelles/surfactant adsorbing inside the microgels preventing them from deswelling. This results in a particle network without the ability to rearrange, resulting in the formation of collapsing gels. The bottom layer highlights the behaviour when samples are heated slowly. At low temperatures the samples are again repulsive, then depletion occurs at intermediate temperatures when the pNIPAM is still swollen. This results initially in the formation of liquid droplets, which are rich in microgel, and eventually in phase separation.}
  \label{fgr:scheme}
\end{figure*}

\begin{figure}
\centering
  \includegraphics[height=2cm]{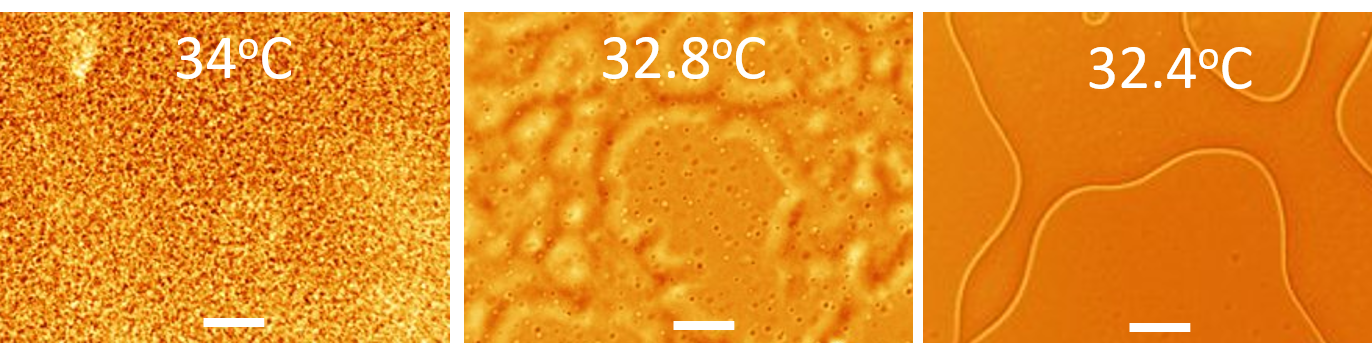}
  \caption{DIC images of a sample of pNIPAM (4 wt\%) and triblock-copolymer (8 wt\%) on cooling. The sample was initially heated at a rate of 10 $^{\circ}$Cmin$^{-1}$ to 35$^{\circ}$C, forming a uniform gel structure. The sample was then cooled at a rate of 0.1$^{\circ}$Cmin$^{-1}$. This was to highlight that the droplet liquid phase can be accessed both on heating and cooling. Scale bar 40 $\mu$m.}
  \label{fgr:dropend2}
\end{figure}

In the case of slow heating, macroscopic observation of the phase behaviour were consistent with the microscopic observations discussed above. When samples at the same composition were heated quickly, by immersion in hot water, all samples formed collapsing gels (Fig. \ref{fgr:phase} a,d). The temperature of the hot water was approximately 70$^{\circ}$C. Here the pNIPAM microgels all aggregate leaving a liquid supernatant consisting of water and excess triblock-copolymer micelles \cite{fussell2019}. When the samples were left in an oven to heat slowly from room temperature to 40$^{\circ}$C, higher concentration samples instead showed evidence of phase separation. A layer rich in pNIPAM formed along with a layer of clear liquid (Fig. \ref{fgr:phase} c). These samples flowed upon inversion of the vial, highlighting that the sample forms a weaker structure than the collapsing gel networks. This difference in structure with heating rate was only observed in samples where demixing was also observed under the DIC microscope. Samples that formed gel networks microscopically on slow heating also formed collapsing gels macroscopically (Fig. \ref{fgr:phase} b). 


\begin{figure*}
\centering
  \includegraphics[height=7cm]{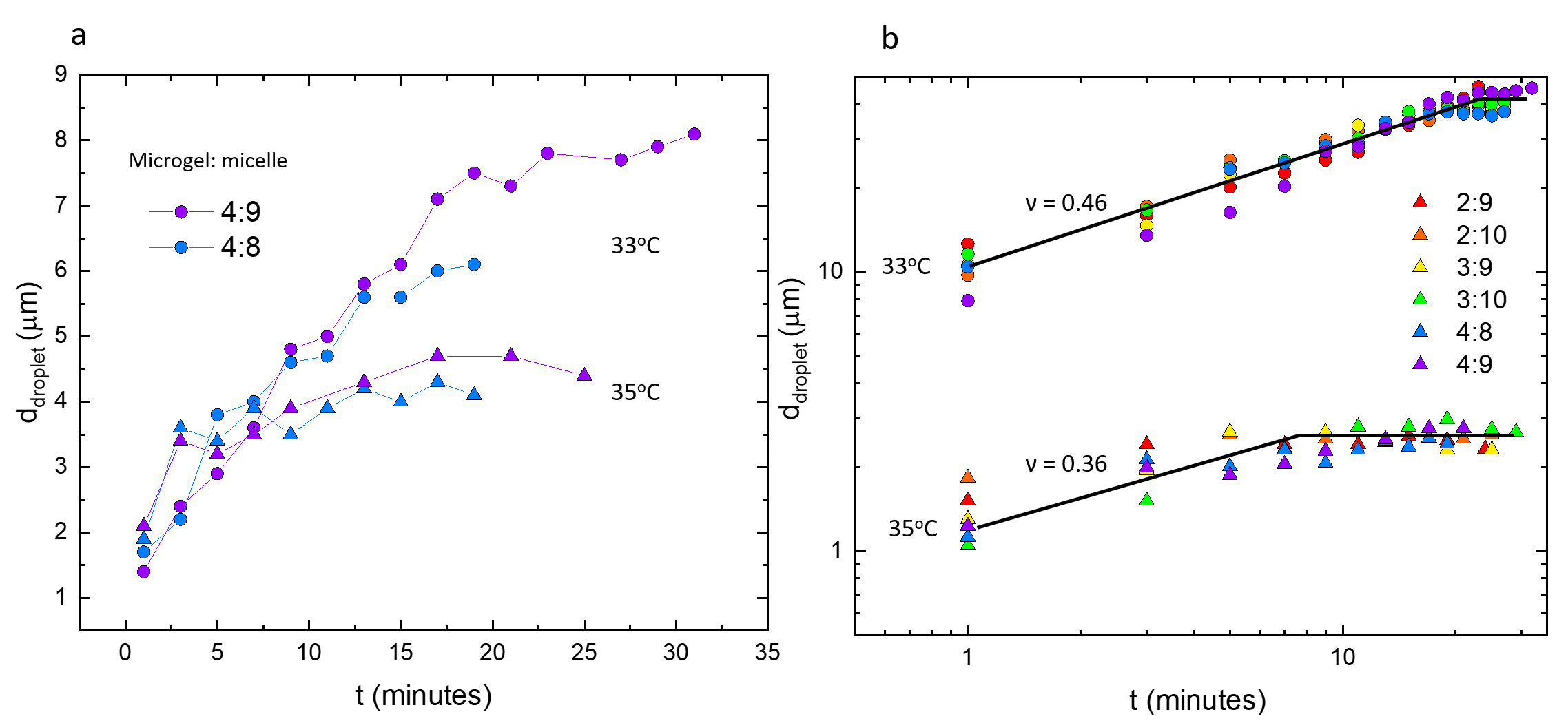}
  \caption{An example of the droplet size evolution over time for two concentrations where droplets were indicated to form in Fig. \ref{fgr:temp2}. (a) Droplet size evolution over time when the samples are heated at a rate of 0.1$^{\circ}$Cmin$^{-1}$, to a final temperature of 33$^{\circ}$C (circles) or 35$^{\circ}$C (triangles).  The time is started from the first indication of aggregates forming. Samples had 4 wt\% pNIPAM, 9 wt\% triblock (purple) and 4 wt\% pNIPAM, 8 wt\% triblock (blue). (b) Shows data for a wider range of concentrations, on a log-log scale. The fitted power law for each droplet series measured is also included. Each data set has been scaled by a pre-factor to highlight the similar trend observed at each concentration. The data at 33$^{\circ}$C has been shifted to separate the data sets on the graph. The black lines are the average fitted power laws for the droplet growth series at both temperatures for all concentrations that formed droplets. The data at 35$^{\circ}$C has been fitted up until the plateau in droplet size. For the concentration, the first number indicates the wt\% of pNIPAM microgels in the sample, and the second number the wt\% of triblock-copolymer micelles.}
  \label{fgr:drop2}
\end{figure*}



\subsection{Aggregation behaviour under slow heating}

In the case of the slow heating rate, we now consider the phase boundary between fluid (low temperature) and gel or demixed phases at higher temperature as a function of concentration of microgels and triblock-copolymer micelles. This is shown in Fig. \ref{fgr:temp2}. We find changes in the nature of the demixing can be obtained as a function of composition as well as heating rate, as discussed above.
At temperatures below approximately 33$^{\circ}$C droplet formation and phase separation occurred  (Fig. \ref{fgr:temp2} a, blue data points), whereas at higher temperatures gels are formed (Fig. \ref{fgr:temp2}, pink data points). 
 
For samples undergoing demixing, the aggregation occurs below the $T_{\text{VPT}}$ for several concentrations of polymer (Fig. \ref{fgr:temp2}). This shows that aggregation in these systems occurs when the particles are still swollen. When gels result (Fig. \ref{fgr:temp2} pink), aggregation occurs when the pNIPAM is close to fully deswollen. However, for the concentrations where the aggregation temperature is lower than the $T_{\text{VPT}}$,  a different phase behaviour is observed and phase separation occurs. The confocal images indicate a change of interaction mechanism with temperature. At lower temperatures the mechanism for aggregation is depletion, where the micelles behave like non-adsorbing polymer resulting in aggregation, evidenced by the micelle signal remaining in the supernatant around the droplets. At higher temperatures than the $T_{\text{VPT}}$, gels result and the confocal signal of the micelles and microgels overlap. 

At room temperature, the microgels are swollen with water, their periphery is poorly defined and the particles are purely repulsive. vdW attractions between the particles are very weak, hence no depletion is observed at these temperatures. At intermediate temperatures, the microgels deswell, becoming more attractive due to increased vdW interactions and more defined due to the periphery collapsing. The micelles cause depletion interactions between these partially swollen particles. The concentrations of micelles needs to be high in order to see depletion interactions as microgels are more stable to depletion than hard spheres \cite{Bergman2018}.
Droplets form at these intermediate temperatures and high concentrations. To form droplets through depletion, the interactions need to be moderately strong and of sufficiently long range. Our micelle / microgel size ratio at these concentrations is approximately 0.25 so is of the order where a colloidal liquid is thermodynamically stable in colloid polymer mixtures \cite{IlettWCKPoonPuseyAOrrockMKSemmlerSErbit}. At longer times, the droplets coalesce and fully phase separate.

At high temperatures the microgels can be modelled as attractive spheres, and depletion interactions no longer occur in these systems as the micelles now associate with the microgels. In the systems where the samples are heated slowly, aggregation due to depletion has already taken place, hence the droplet structures and phase separation remain once heated to the temperature where association occurs. 

The temperature of the phase boundary decreases with increasing micelle concentration, as shown in Fig. \ref{fgr:temp2}(a). The phase boundaries are shown for various microgel concentrations from 1--6 wt\%.  At low concentration of microgels (1 wt\%), we encounter only gelation. Increasing the concentration of microgels to 6 wt\%, we find phase separation at a progressively lower micelle concentration. Also at low concentration of microgels, the phase boundary is almost flat as a function of micelle concentration, but slopes down progressively upon increasing the microgel concentration. 

In Fig. \ref{fgr:temp2}(b) we perform the complementary analysis of holding the micelle concentration fixed, and plot the temperature of the phase boundary as a function of microgel concentration. We find a very similar behaviour, in that we find phase separation at a progressively lower micelle concentration. Furthermore, the phase boundary is again almost flat at low micelle concentration but slopes down more and more steeply upon increasing the micelle concentration. This suggests that the phase behaviour may depend on the \emph{combined} concentration of micelles and microgels. To test this, in Fig. \ref{fgr:temp2}(c) we plot the phase boundary as a function of the combined concentration of micelles and microgels. Remarkably, we find a collapse of all data to a small spread around a single curve. This is a \emph{highly surprising} result, because it suggests that we can treat the micelle and microgel concentrations interchangeably, even though these are wildly different species.

\subsection{Mechanistic description of phase behaviour}
\label{sectionMechanisticDescription}

The interaction potential between pure pNIPAM microgels changes as the particles collapse. At low temperatures, microgels are soft colloidal particles, with a layer of extended linear polymer chains. These swollen particles are purely repulsive, vdW interactions are very small, due to the highly diffuse nature of the outer regions of the particles \cite{Bergman2018}. When the microgels are heated, they collapse and are often modelled as attractive spheres, due to the increased vdW interactions and the microgels becoming increasingly hydrophobic \cite{Bergman2018}. The collapsed particles are more likely to aggregate due to these attractions. These colloids are stabilised solely by the charge on the particles left from the initiator during synthesis.

We now consider mixtures of microgel particles and the triblock-copolymer micelles, as summarised in Fig. \ref{fgr:interactions}. At low temperature, the interaction potential between microgels is soft and purely repulsive, as inferred from DLS measurements which showed no sign of aggregation \cite{fussell2019}, as indicated in Fig. \ref{fgr:interactions}(a), therefore fluid phases are observed. Due to the weak interactions between microgels and their soft nature they remain repulsive even in the presence of the depletant. The micelles are also temperature responsive, and at low temperatures contain high volume fractions of water, therefore act as weak depletants at room temperature. When the micelles are heated, it was found that the hard sphere volume fraction at equivalent concentrations increased \cite{Sharp2007}.

At intermediate temperatures, the interaction potential becomes increasingly attractive as the microgels collapse, where vdW attractions between the microgels start to appear close to the $T_{\text{VPT}}$. Due to the microgels still being partially swollen at these temperatures, the vdW potential well is relatively shallow and the interaction is relatively soft (Fig. \ref{fgr:interactions} (b)). The phase separation at these temperatures is driven by depletion, where the non-adsorbing micelles cause aggregation. At these concentrations and temperatures, the depletion interaction results in droplet formation and with increased temperature and time, phase separation. Due to microgels being more stable to depletion compared to hard spheres at low temperatures \cite{Bergman2018}, this droplet phase is only observed at the highest concentration of micelles (depletant). At these intermediate temperatures, there was no evidence of association using DLS, evidenced by the hydrodynamic diameter of the microgels at these temperatures being similar to the pure microgel even in the presence of triblock-copolymer  \cite{fussell2019}. At these intermediate temperatures, it has been observed using small angle neutron scattering that the aggregation number, volume fraction of polymer in the micelle and hard sphere volume fraction all increase with temperature, while to core radius of the micelles remains constant \cite{Sharp2007}. As the volume fraction of micelles increases, so does the volume fraction of depletants, hence the depletion interaction is stronger at intermediate temperatures.

When the temperature is increased further, the microgels fully collapse. If droplets have already formed they remain with increased temperature, however they appear to coarsen. If no depletion induced phase separation has taken place yet, at low micelle concentrations, the microgels and the micelles associate. These microgel particles are likely prevented from deswelling by the micelles, increasing the vdW interactions between the particles inducing gelation. The microgels are harder, and the vdW forces are stronger, corresponding to a deeper attractive potential well (Fig. \ref{fgr:interactions} c). This results in the particles being unable to rearrange once association occurs, resulting in collapsing gels. Figure \ref{fgr:scheme} (lower scheme) captures how the interaction potentials relate to the structures observed. 

The difference in behaviour with heating rate indicates that the equilibrium phase is not accessed when the samples are heated quickly (10$^{\circ}$Cmin$^{-1}$), resulting in gelation. In this regime, the microgels and the micelles associate very quickly as the temperature increases rapidly above that at which the system is dominated by microgel--micelle attractions, and the collapse of the microgels also occurs over a fast timescale. This results in the depletion effects at intermediate temperatures being bypassed, resulting in gel network formation (Fig. \ref{fgr:scheme} top scheme). Due to the strong vdW forces between the particles, once the network has formed, there is little ability for particles to rearrange. Therefore the kinetically favoured phase forms: collapsing gel networks that undergo syneresis. 

\subsection{Detailed phase behaviour}
\label{sectionDetailedPhaseBehaviour}

So far, we have discussed the overall behaviour of the system at fast and slow heating rates and have identified a mechanism. We now proceed to examine in more detail certain phenomena and characteristics of our system. In particular, we consider the detailed structure of the gel, the droplet phase evolution and reversibility.

\emph{Detailed structure of gels. ---}
Figure \ref{fgr:phasediagram} details the phase behaviour observed, when the samples were heated slowly (0.1$^{\circ}$C\ min$^{-1}$). For low concentrations of micelle and microgel, the system gels. Two types of networks were observed. The first was uniform gels, where little structural detail was observed at the resolution of the microscope used (Fig. \ref{fgr:phasediagram} dark pink). These were formed at high concentration of microgels, more than 3 wt$\%$. The second morphology we term gels with voids, where, when imaged under the DIC  microscope the arms of the gel network were resolvable (Fig. \ref{fgr:phasediagram} light pink). At the lower concentrations of microgel $(\lesssim 3$wt$\%)$, gels with voids preferentially form, as the low volume fraction of microgel in these samples results in sparse networks. These low concentration samples macroscopically do not form gels, but only show evidence of aggregates which sediment \cite{fussell2019}. 

Turning to the behaviour of the system as a function of micelle concentration, we find that gels with voids also form preferentially with increasing micelle concentration. This is most prevalent at concentrations close to the boundary where phase separation is observed. To explain this trend we looked in detail at the aggregation temperature as a function of both micelle and microgel concentration. This is summarised in Fig. \ref{fgr:temp2} (b and c). It can be seen that the aggregation temperature decreases with increasing micelle concentration. This decrease in aggregation temperature means that aggregation occurs when the particles are partially swollen. Now, at these lower temperatures, the interaction between the micelles and microgels is one of depletion (see Section \ref{sectionMechanisticDescription} and Fig. \ref{fgr:interactions}). This results in phase separation occurring before gelation, as the microgels are free to rearrange to some extent over short time scales before full gelation occurs, resulting in these voids.

At the higher concentrations of microgels and micelles, demixed phases were observed. At these concentrations samples form microgel-rich phases (Fig. \ref{fgr:phasediagram}). At lower microgel and micelle concentrations these droplets remain stable over the time frame of the experiment (light blue), but over long time scales (hours) they eventually coalesce leading to macroscopic phase separation (ESI Fig. S3). With increasing polymer concentration the rate of droplet coalescence increases, and phase separation proceeds (Fig. \ref{fgr:phasediagram} dark blue) over the time frame of the experiment. 

At higher microgel concentrations, lower micelle concentrations are needed to induce phase separation. At micelle concentrations above $\sim7$ wt\%, the aggregation temperature becomes strongly dependent on microgel concentration and drops rapidly (Fig. \ref{fgr:temp2} b, c). Phase separation (droplets coalescing) was found to occur when the aggregation temperature was lowest (below 31$^{\circ}$C), as aggregation in these samples occurs over the largest time / temperature range.


\subsection{Reversible phase behaviour}

The formation of the droplet phase appears to be reversible, highlighting that the phase is stable. However it is found over a small temperature range only, between the temperature at which aggregation occurs and where the microgels are fully deswollen. We took a mixture of microgels and micelles known to form droplets when heated slowly (4 wt\% pNIPAM and 8 wt\% triblock-copolymer). We heated this sample rapidly to 35$^{\circ}$C so the sample gelled. The sample was then cooled at at rate of 0.1$^{\circ}$C\ min$^{-1}$. On cooling the droplet phase was again observed, eventually phase separating into polymer rich and poor liquid phases. This shows that the droplet phase is stable as it can be accessed both on heating and cooling the system.

\subsection{Droplets of Colloidal Liquid}

We have characterised the droplets of colloidal liquid (microgels) that result from the slow heating of our system. Droplets only form at high micelle concentrations and temperatures below the $T_{\text{VPT}}$. The size of the droplets increases with time, tending to a plateau value as shown in Fig. \ref{fgr:drop2}(b). We believe that the suppression of full phase separation and the trend towards a plateau is due to the stronger interactions found between the microgels at temperatures of 35$^\circ$C or more. 

The size of the droplets formed can be controlled by changing the temperature the droplets are held at. Two different experimental regimes were studied as shown in Fig. \ref{fgr:drop2}. The first of these being where samples were heated at 0.1$^{\circ}$Cmin$^{-1}$ from room temperature to 33$^{\circ}$C, which is just above the temperature where aggregation was observed for these samples. The second was where samples were heated from room temperature to 35$^{\circ}$C also at a rate of 0.1$^{\circ}$Cmin$^{-1}$, a temperature where pNIPAM is nearly fully deswollen (Fig. S2 ESI). Two example data sets have been included in Fig. \ref{fgr:drop2}, the data for all phases where droplets occur is included in the SI (Fig. S4 ESI). For all samples, time zero was the time after the temperature at which aggregates were first observed. An example set of images of the droplets increasing in size over time is included in the SI (Fig. S5 ESI). The average difference in the diameter of the droplets is 48\%$\pm$2\% between 33$^{\circ}$C and 35$^{\circ}$C. The microgels themselves deswell approximately 39\% of their overall diameter over this temperature range. It is likely that at the lower temperature of 33$^{\circ}$C, the droplets can continue to grow as the interaction between the microgels are relatively weak at this temperature, due to the particles being partially swollen, so the polymers are free to rearrange resulting in droplet growth for longer times. However when the samples are heated to 35$^{\circ}$C, the association between the microgels becomes relatively strong, meaning the microgels are no longer free to rearrange because of the stronger attractions between them. 

The evolution of the droplet diameter with time was fitted to a power law $y=Ax^{\nu}$ for both temperature regimes (33$^{\circ}$C and 35$^{\circ}$C). The data sets were scaled by a prefactor A to highlight the constant power law for each temperature. The data at 33$^{\circ}$C has been shifted to higher values to separate the data sets (Fig. \ref{fgr:drop2}b). Each data set was fitted up to the point where the droplet growth plateaus. Both final temperatures showed a power law regime for droplet size growth as a function of time, where a faster evolution was revealed for 33$^{\circ}$C than for 35$^{\circ}$C. This is to be expected as the droplets grew to a larger size at the lower temperature over the same time frame. At 33$^{\circ}$C $\nu$=0.46 and at 35$^{\circ}$C $\nu$=0.36. There are two important factors to consider for droplet evolution, diffusion and coalescence. At low droplet concentrations, there is a negligible interaction between the droplets, this is the diffusion limited case, $\nu$ = 1/3. When the concentration of droplets increases, droplet coalescence becomes an important factor, $\nu$ increases towards 1 \cite{Steyer1991}. From the values of the exponents that we determine, it seems that with increasing temperature, droplet coalescence becomes less important, and the droplet evolution is driven by diffusion (35$^{\circ}$C). At lower temperatures, droplet coalescence becomes more important (33$^{\circ}$C) hence the increased $\nu$. This again will be due to the limited ability of the microgels to reorganise at higher temperatures due to the stronger attraction between them. At lower temperatures the microgels are still swollen, resulting in a weaker interaction compared to the collapsed particles resulting in droplet growth occurring over longer time scales.

The size of the droplets observed correlates with the concentration of both microgels and micelles in the samples. With increasing microgel concentration, there is a linear increase in the size of the droplets observed (Fig. S6 ESI). A similar trend is observed for the micelle concentration (Fig. S7 ESI). Both of these trends are observed when the samples are heated to 33$^{\circ}$C and 35$^{\circ}$C.

\section{Discussion and conclusions} \label{sectionResults}

We have identified a colloidal model system displaying tunable, controllable gelation, building on the current understanding of the phase behaviour the PEO-PPO-PEO triblock copolymer pNIPAM microgel system, in which the former form micelles, leading to a micelle-microgel mixture. Our study reveals a previously unreported phase separation to a phase rich in microgels coexisting with a rich in micelles and poor in microgels. While previous work \cite{fussell2019} looked at rapid temperature changes, here we have explored the effect of heating rate, which, along with the composition, has revealed a range of structures formed in this non-equilibrium system. Demixing in our system at low temperatures is driven by depletion interactions, which are negligible at room temperature, but become apparent with increased temperature due to the increased ability of the micelles to be depleted at higher temperatures. At high temperatures, an associative mechanism of gelation dominates, as hydrophobic interactions and hydrogen bonding cause the microgels and micelles to associate.

These mixtures from temperature responsive gels upon heating, and we have shown that the structure and phase of these systems can be controlled by altering the concentration of the components and by altering the heating rate used for gelation. We highlight the broad range of states accessible for mixtures of pNIPAM microgels and triblock-copolymer micelles. In particular, upon high rates of heating, phase separation is suppressed, and gelation is encountered. At lower rates of heating, the system spends sufficient time at temperatures where depletion dominates and the attractions between the microgels are milder, and a more complete demixing occurs, with the formation of liquid--like droplets of microgels. This is reminiscent of the coupling between the rate of change of attraction in ``sticky spheres'' and the degree of self-assembly \cite{Royall2012}.

In addition to its response to the heating rate, our system can be tuned by changing the composition. High concentrations of microgel particles lead to uniform gels, while lower concentrations result in gels with voids. Increasing the concentration of micelles promotes demixing to droplets of microgels and micelles (which then solidify). At higher micelle concentration still, we encounter full phase separation to a phase rich in microgels in coexistence with a microgel poor phase. The temperature at which the microgels demix is dependent on the concentration of both the microgels and the micelles, and at high concentrations demixing occurs when microgels are still in the swollen state. 

This system highlights how studying the phase behaviour of a microgel-micelle system can ultimately improve our understanding of the behaviour of colloidal assemblies and for example can allow us to access both the equilibrium phase behaviour and non-equilibrium gel and droplet structures. Furthermore these systems may find use as novel temperature-responsive materials.
\section*{Acknowledgements} \label{sec:acknowledgements}

SF is supported by a studentship provided by the Bristol Centre for Functional Nanomaterials (EPSRC grant EP/L016648/1). CPR acknowledges support from EPSRC (grants GR/M32320/01 and ERC consolidator grant NANOPRS, project number 617266). We would also like to thank Craig Davies, Croda, for supplying the Synperonic PE/P105. 

\bibliography{references}
\bibliographystyle{rsc}

\counterwithin{figure}{section}

\section{Appendix} \label{sec:appendix}

\renewcommand{\thefigure}{S\arabic{figure}}

Figure S1 highlights how differential interference contrast (DIC) images were edited in this work. The images were converted to grey scale, then converted to the glow colour palette, then the image inverted. The brightness of the image was then adjusted. This results in the polymer rich regions appearing bright and the polymer poor regions appearing dark. 

\begin{figure}[h]
\centering
  \includegraphics[width=\columnwidth]{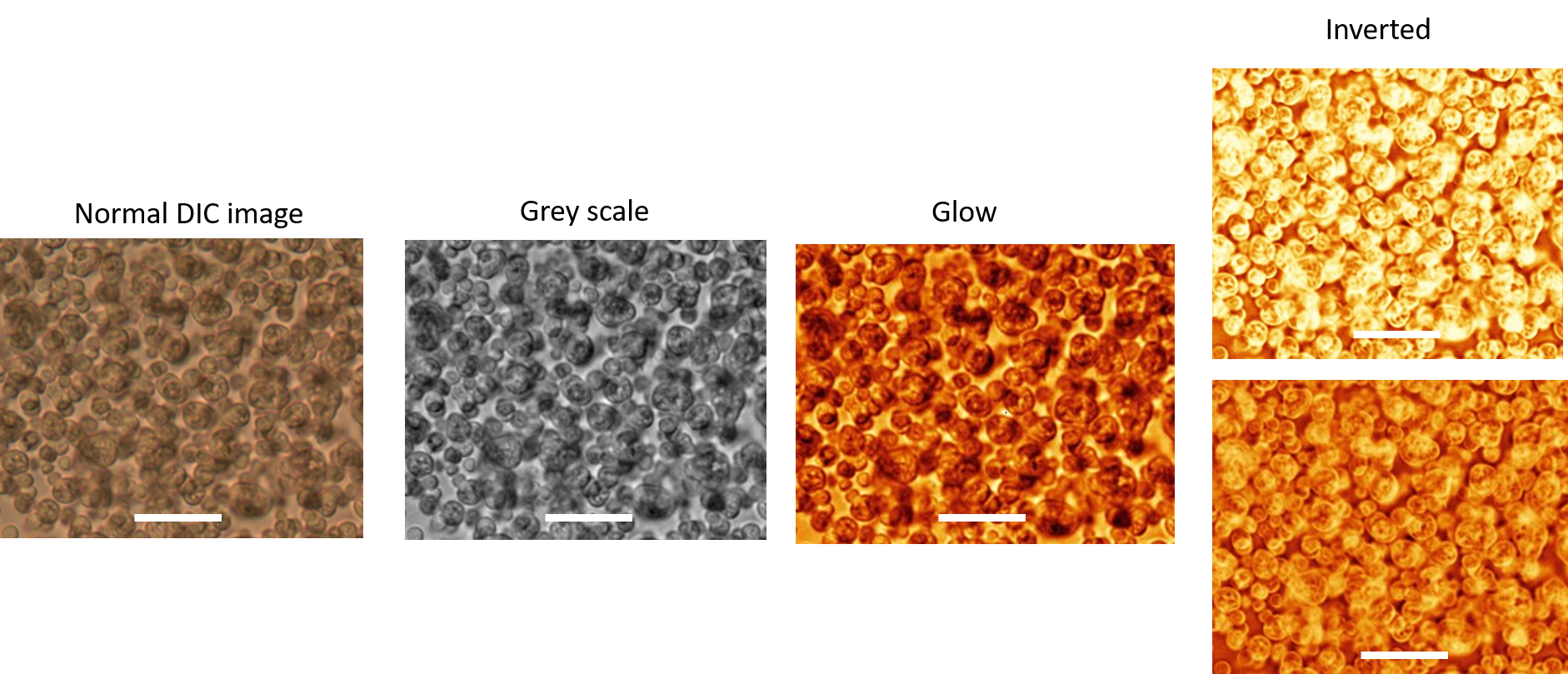}
  \caption{Figure highlighting the editing process used to highlight the regions rich in polymer in images taken using the DIC microscope. DIC images are converted to grey scale, then converted to glow, then the images inverted so that the polymer rich regions appear bright and the polymer poor regions appear dark.}
  \label{fgr:SIUCL}
\end{figure}

\begin{figure}
\centering
  \includegraphics[width=\columnwidth]{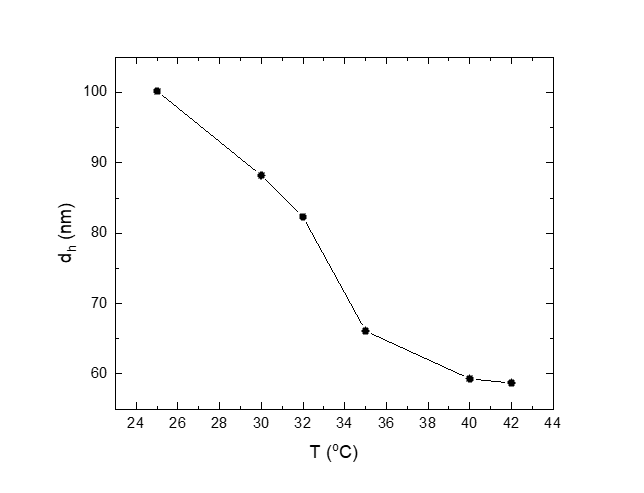}
  \caption{Hydrodynamic diameter of the pNIPAM microgels as a function of temperature.}
  \label{fgr:SIUCL}
\end{figure}

\begin{figure}[h]
\centering
  \includegraphics[width=\columnwidth]{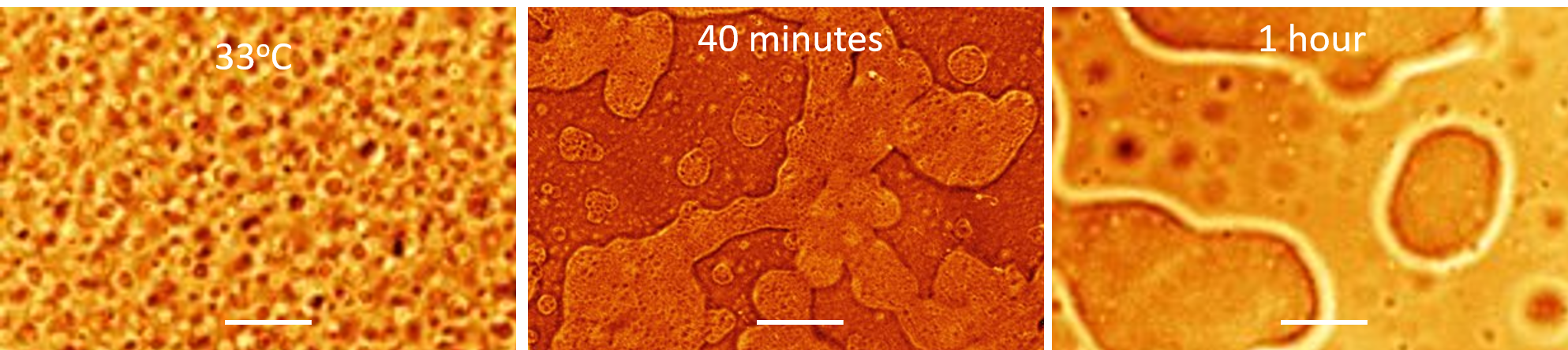}
  \caption{Images of 
  droplets coalescing over time. The concentration of the sample is 4 wt\% pNIPAM and 8 wt\% triblock-copolymer. The scale bar is 40 $\mu$m. }
  \label{fgr:dropend}
\end{figure}

\begin{figure}[h]
\centering
  \includegraphics[height=7cm]{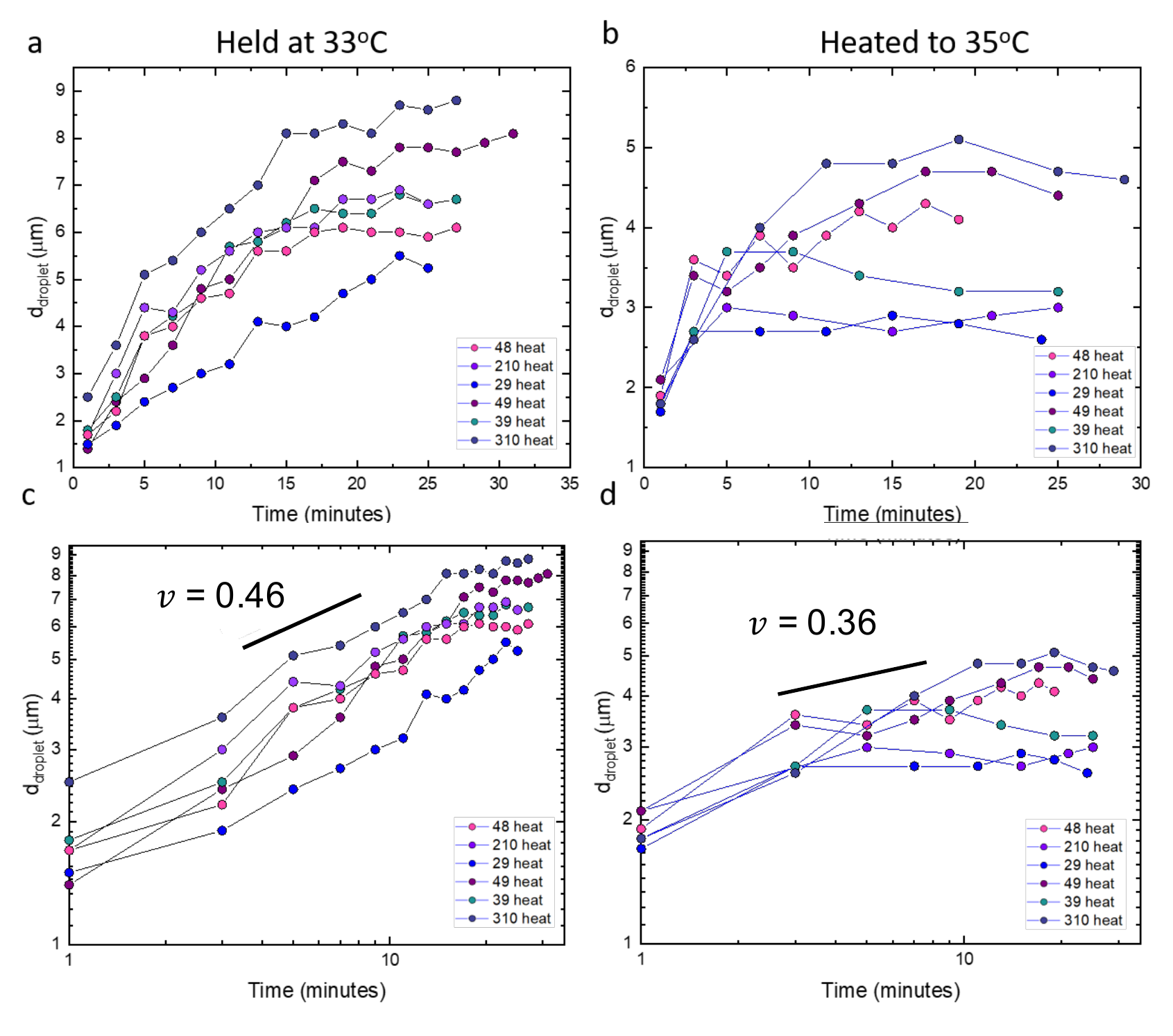}
  \caption{Summary of the droplet size evolution over time, for the concentrations where droplets were indicated to form in figure 1. Droplet size evolution over time when the samples are heated at a rate of 0.1 $^{\circ}$C/min to 33$^{\circ}$C (a) or 35$^{\circ}$C (b). The same data has been plotted on a log time scale in (c) and (d). The time is started from the first indication of aggregates forming. 4 wt\% pNIPAM, 9 wt\% triblock (maroon), 4 wt\% pNIPAM, 8 wt\% triblock (pink), 3 wt\% pNIPAM, 10 wt\% triblock (grey), 3 wt\% pNIPAM, 9 wt\% triblock (green), 2 wt\% pNIPAM, 10 wt\% triblock (purple), 2 wt\% pNIPAM, 9 wt\% triblock (blue) The black lines are the average calculated fitted power laws for the droplet growth series at both temperatures for all droplet sizes in the series.}
  \label{fgr:scheme}
\end{figure}

Figure S2 contains the deswelling data for the pNIPAM microgels used in this study, determined using dynamic light scattering. 

Figure S3 highlights the evolution of the droplet phase over time. Droplets begin to form upon heating, then when held at 33$^{\circ}$C, the droplets continue to grow and eventually merge. This results in phase separation at prolonged times. 

\begin{figure}[t]
\centering
  \includegraphics[height=5cm]{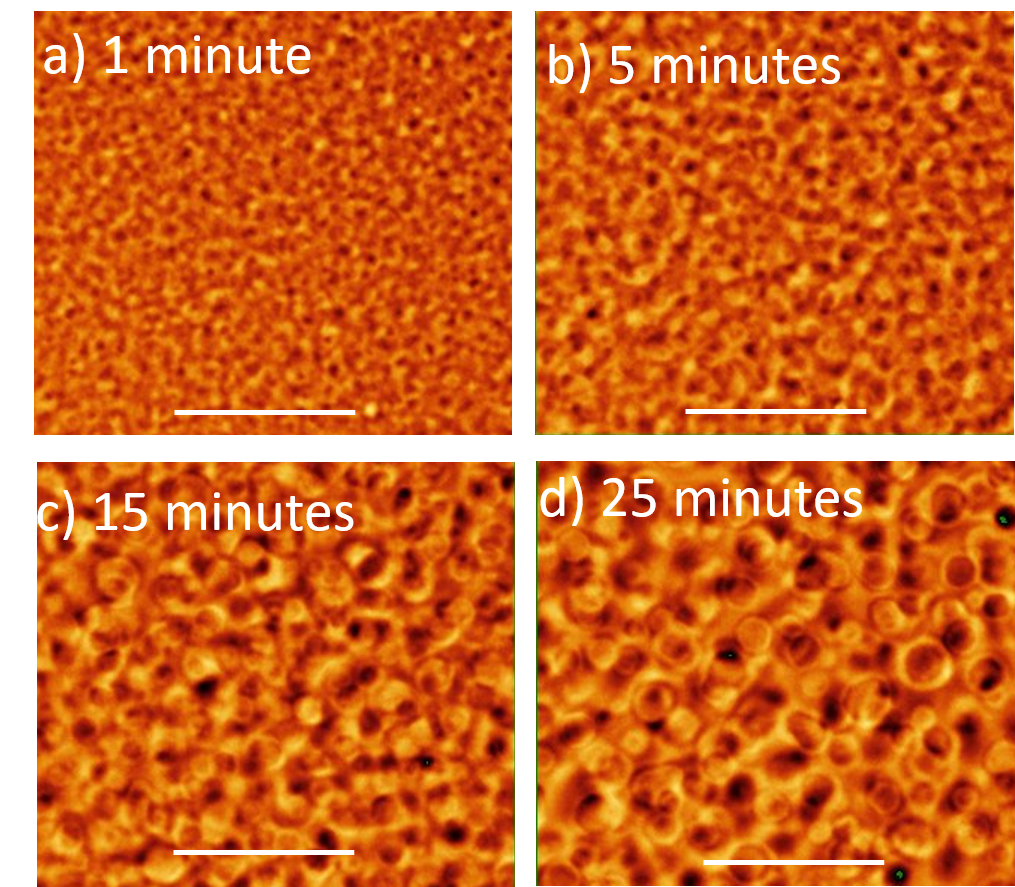}
  \caption{a-d) Optical micrographs of 3 wt\% pNIPAM, 9 wt\% triblock-copolymer heated to 33$^{\circ}$C at a rate of 0.1 $^{\circ}$C/min and held at that temperature. The time was started from the first indication of gelation. The scale bar is 10 $\mu$m}.
  \label{fgr:drop}
\end{figure}

\begin{figure}[h]
\centering
 \includegraphics[width=\columnwidth]{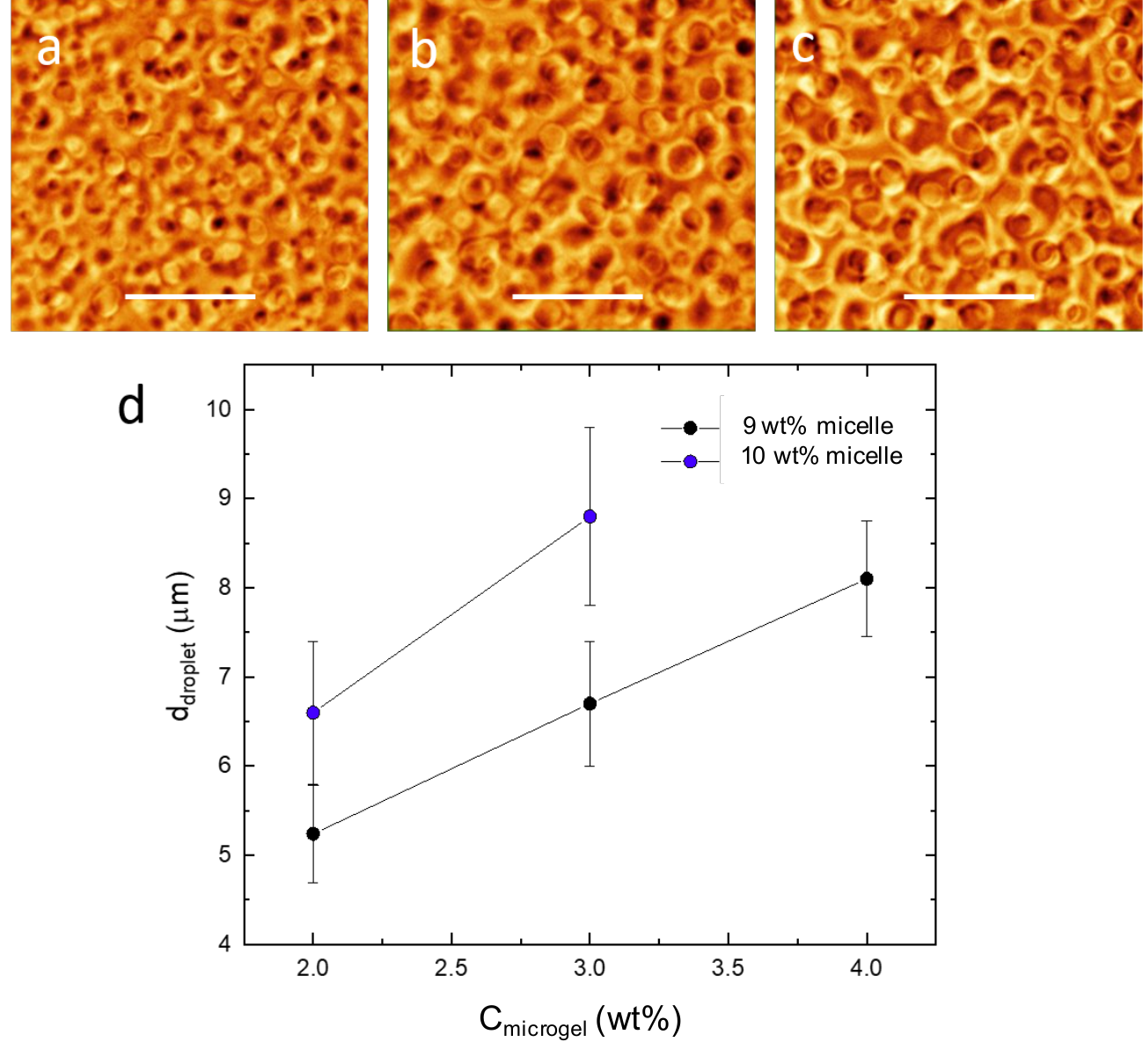}
    \caption{Summary of how droplet size increases with increasing pNIPAM concentration, for samples with 9 wt\% triblock-copolymer. a) 2 wt\% pNIPAM, b) 3 wt\% pNIPAM, c) 4 wt\% pNIPAM. Scale bar 40 $\mu$m. d) Summary of the final droplet sizes recorded when samples were heated to 33$^{\circ}$C at a rate of 0.1 $^{\circ}$C/minute and held at that temperature.}
  \label{fgr:dropend2}
\end{figure}

Figure S4 shows the evolution of droplet size for the series of samples that form droplets. There is a general increase of droplet size with time, tending to a plateau value at late time. It can be seen that droplets grow to larger sizes when held at 33$^{\circ}$C (a), rather than heated to 35$^{\circ}$C  and held at that temperature. It was also observed that a power law can be used to describe the data set, where samples heated to 33$^{\circ}$C scale with gradient $\mu$ = 0.46, whereas samples heated to 35$^{\circ}$C scale with $\mu$ = 0.36.


\begin{figure}[h]
\centering
  \includegraphics[width=\columnwidth]{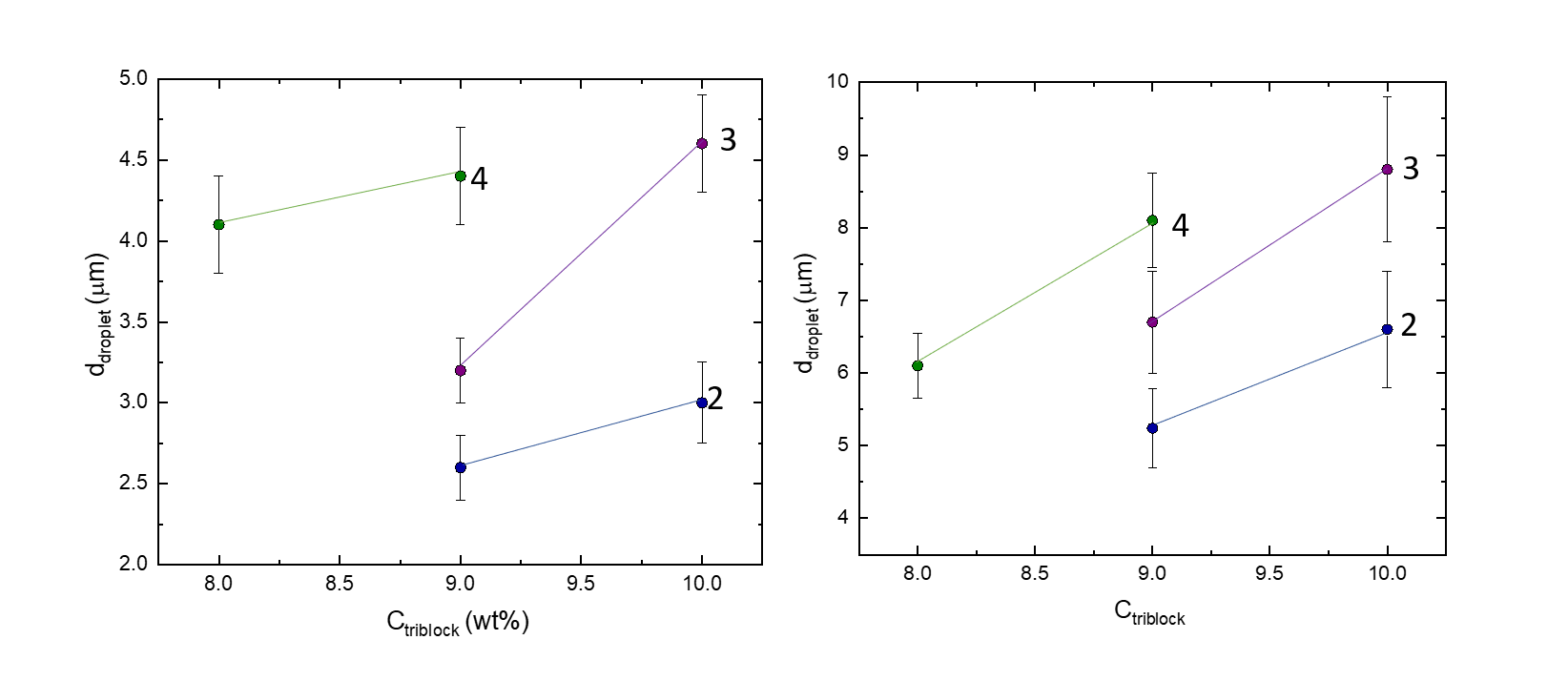}
  \caption{Summary of the final droplet sizes observed from heating mixtures of pNIPAM and triblock-copolymer. Left are the final droplet sizes observed for samples being heated at a rate of 0.1 $^{\circ}$C/min to 35$^{\circ}$C. Right are for samples heated to 33$^{\circ}$C at a rate of 0.1 $^{\circ}$C/min. The numbers on the graph indicate the concentration of pNIPAM in wt\%. Coloured lines have been added to guide the eye.}
  \label{fgr:dropend}
\end{figure}

Figure S5 illustrates how the size of the droplets increases with time. 

Figure S6 and S7 shows how the size of droplets, formed through slowly heating mixtures of pNIPAM and triblock-copolymer, changes as a function of pNIPAM and triblock-copolymer concentration respectively. With increasing polymer concentration there is a general increase to the size of the droplets observed. This effect was seen for various pNIPAM concentrations and for different final temperatures where the drops were held.

\end{document}